\documentclass[%
 aps,
 prb,%
 amsmath,amssymb,
 preprint,%
 notitlepage,
]{revtex4-2}
\usepackage[T1]{fontenc}
\usepackage{graphicx}
\usepackage{dcolumn}
\usepackage{bm}
\usepackage{appendix}
\usepackage{amsmath}
\usepackage{amssymb}
\usepackage{graphicx}
\usepackage{indentfirst}
\usepackage{booktabs}
\usepackage{multirow}
\usepackage{colortbl}
\usepackage{hyperref}
\begin{document}


\title{Dissecting Exciton-Polariton Transport in Organic Molecular Crystals: Emerging Conductivity Assisted by Intermolecular Vibrational Coupling} 

\author{Guangming Liu}
\affiliation{Department of Chemistry and Biochemistry, University of Notre Dame, Notre Dame, IN, 46556}
\author{Hsing-Ta Chen}
\affiliation{Department of Chemistry and Biochemistry, University of Notre Dame, Notre Dame, IN, 46556}

\begin{abstract}
In this work, we systematically investigate the spectral and transport properties of exciton-polaritons under the explicit influence of intermolecular vibrational coupling, which introduces dynamic disorder. In the context of a one-dimensional molecular chain strongly interacting with a cavity photon, we demonstrate the polaritonic characteristics of the spectral function and its interactions with the electronic band broadened by the coupling disorder. 
We further dissect the current flux into its bare excitonic contribution and transport via the cavity photon. 
Our results reveal that the enhancement in the charge carrier mobility and frequency-resolved conductivity stems from the photon-mediated current. 
More importantly, contrary to the intuition that dynamic disorder hinders transport, intermolecular vibrational coupling can facilitate exciton-polariton transport, offering an additonal degree of tunability for material properties.
\end{abstract}{}
%




\maketitle


\section{Introduction}
Harnessing strong light-matter interactions offers a promising strategy for modifying charge carrier transport of molecular systems confined within an optical microcavity or integrated with a plasmonic nanostructure\cite{xu_ultrafast_2023,balasubrahmaniyam_enhanced_2023,sandik_cavity-enhanced_2025,jin_enhanced_2023,berghuis_controlling_2022,wurdack_motional_2021,bhatt_enhanced_2021,park_polariton_2022,chen_unraveling_2023}.
On the one hand, the collective and delocalized nature of exciton-polariton, a hybrid quasi-particle arising from the strong coupling of the radiative molecular excitations and the confined photonic fields, points towards enhancement in long-range and ultra-fast transport\cite{feist_extraordinary_2015,schachenmayer_cavity-enhanced_2015}.
On the other hand, the inherent disorder stemming from lattice distortion and intermolecular interactions causes random scattering of charge carriers, leading to relaxation and localization\cite{coles2011vibrationally,engelhardt_polariton_2023,catuto_interplay_2025}, thereby suppressing the overall exciton-polariton transport.
Despite these intuitive effects, however, recent studies on exciton-polaritons of disordered molecular systems show that disorder effects can induce transport enhancements as the dark states (molecular excitations that are not coupled to photons) can acquire long-range transport characteristics of polaritons\cite{aroeira_coherent_2024,tutunnikov_characterization_2024,suyabatmaz_vibrational_2023,ribeiro_multimode_2022,allard_disorder-enhanced_2022,tichauer2022identifying}.
Thus, understanding the microscopic interplay between exciton-polariton and molecular disorder has posed a significant theoretical challenge for designing efficient functional materials.

Among many disordered molecular systems, organic molecular crystals play crucial roles in photovoltaics, display technologies, plastic electronics, and spintronic devices\cite{forrest2004path,muccini2006bright,sirringhaus201425th,bredas2004charge,bredas2009molecular}. 
In contrast to covalently bonded crystalline inorganic materials, organic molecular crystals are strongly influenced by two forms of electron-phonon coupling.
First, the intramolecular electron-vibration coupling forms the Holstein-type small polaron, which renormalizes the electronic excitation energy and localizes the charge carrier exciton\cite{holstein_studies_1959,ortmann_theory_2009}. This Holstein-type phonon is usually of high frequency and can be treated using the Lang-Firsov polaron transformation\cite{fetherolf2020unification}.
Second, the interaction between electrons and nonlocal intermolecular vibrations gives rise to dynamic disorder or transient localization\cite{troisi2006charge,wang_mixed_2011,ciuchi2011transient,fratini2016transient}, which leads to delocalization of the electronic wavefunction and eventually influences charge carrier mobility. These nonlocal phonon modes are often of low frequency and can be modeled by Peierls-type (or Su-Schrieffer-Heeger) mechanism\cite{su1979solitons,sumi1979theory}.
In a recent study, Fetherolf \emph{et~al.} employ the Kubo formalism\cite{fan_linear_2021} and investigate the spectral and transport properties of 1D molecular chain in the presence of both forms of electron-phonon coupling\cite{fetherolf2020unification}.
Here, the spectral functions show phonon-dressed electronic bands broadened by dynamic disorder, and the transport properties are usually characterized by the frequency-resolved optical conductivity $\text{Re}[\sigma(\omega)]$ and the charge carrier mobility $\mu$\cite{fetherolf2020unification,bhattacharyya_anomalous_2024}.

In the presence of strong light-matter coupling, the delocalized characteristics of polaritons also significantly affect the microscopic transport properties\cite{hagenmuller2017cavity, chavez2021disorder, bhatt2021enhanced}.
Exciton polariton transport inside an optical cavity is often characterized by different approaches.
To have a direct connection to experimental observations, Mandal \emph{et~al.} have developed microscopic simulations of molecular excitons interacting with multiple photon modes to elucidate multi-mode polariton dispersion relation, featuring the upper and lower polariton bands as a function of the cavity photon wavevector\cite{mandal2023microscopic,mandal_theoretical_2023}.
With the dispersion of polariton bands, one can extract the group velocity of exciton polaritons from the derivative of the dispersion relation\cite{chng_quantum_2025,ying_microscopic_2025,fowler-wright_mapping_2025}.
More explicitly, the propagation of the electronic wavefunction can be obtained by real-space semiclassical electrodynamic simulations\cite{zhou_nature_2024,dini_nonlinear_2024,castagnola_strong_2024}, revealing the correlation between exciton transport and the localized electromagnetic field inside the cavity.
When employing a non-equilibrium protocol, charge carrier mobility can be determined by the time-dependent mean square displacement (MSD) of the molecular excitation wavefunction\cite{krupp_quantum_2024,sokolovskii_multi-scale_2023,sokolovskii_one_2024,koshkaki_exciton-polariton_2025}, in which the characteristic power-law scaling of the MSD reveals whether the expansion is diffusive or ballistic.
Despite many recent progress, there are significant gaps that persist in our understanding of exciton polariton transport within organic molecular crystals\cite{khazanov_embrace_2023,pandya_tuning_2022,pandya_microcavity-like_2021,hou_ultralong-range_2020,rozenman_long-range_2018}. In particular, since existing models mostly consider non-interacting molecules rather than electronic bands of crystalline molecules, it is necessary to clarify how the mobility and optical conductivity of exciton-polariton are impacted by electron-phonon coupling.

In this work, we employ the linear-response Kubo formalism to investigate exciton-polariton transport properties of a molecular chain strongly coupled to a cavity photon mode.  
Notably, in contrast to non-interacting molecules coupled to multiple photon modes, we consider interacting molecules forming an electronic band and focus on the impact of a single photon mode. 
We model dynamic disorder via the Peierls-type electron-phonon coupling and calculate the current autocorrelation functions through quasiclassical dynamics simulations.
This approach allows us to dissect the total conductivity into contributions from delocalized, long-range transport mediated by the cavity photon and that arising from intrinsic intermolecular coupling.
This paper is organized as follows. In Sec.~\ref{sec:theory}, we formulate the system Hamiltonian and the transport properties in the Kubo formalism.  To gain an intuitive picture, we analytically investigate the polariton effect on the transport properties without dynamic disorder in Sec.~\ref{sec:Analysis}. 
We then numerically calculate the spectral function, charge carrier mobility, and optical conductivity using quasi-classical simulation in Sec.~\ref{sec:results}. Finally, we conclude and discuss future directions in Sec.~\ref{sec:conclude}.

\section{Theory}\label{sec:theory}
\subsection{Model Hamiltonian}
\begin{figure}
  \includegraphics[width=\columnwidth]{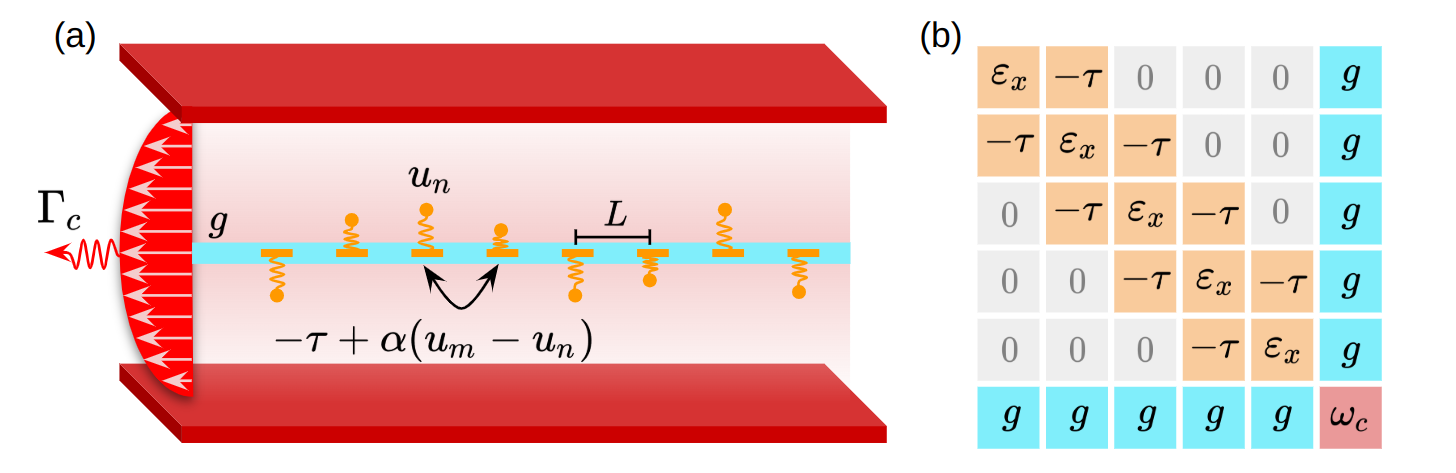}
  \caption{(a) Schematic illustration of a 1D molecular chain inside an optical cavity. $u_n$ denotes the displacement associated with the lattice distortion of site $n$. The exciton transport between the nearest-neighboring molecules is modeled by the bare transfer integral $\tau$ and the Peierls-type electron-phonon coupling $\alpha(u_{m}-u_{n})$. $L$ is the lattice constant. The cavity mode is coupled to the molecular chain with a constant coupling strength $g$, and the decay rate is $\Gamma_c$. (b) The corresponding single excitation matrix in the basis state $|X_n 0\rangle$ for $n=1,\cdots$ and $|G \alpha\rangle$. The orange blocks denote $N$ exciton states with energy $\varepsilon_x$ (diagonal) and the nearest neighbor coupling $-\tau$ (off-diagonal). The cavity photon of frequency $\omega_c$ (red) is coupled to each exciton with a constant coupling strength $g$ (cyan).}\label{fig:schematic}
\end{figure}

To investigate transport properties of organic molecular crystals, we consider a spinless one-dimensional molecular chain of $N$ lattice sites inside an optical cavity as illustrated in Fig.~\ref{fig:schematic}(a) The total Hamiltonian can be described by $\hat{H} = \hat{H}_\text{el} + H_\text{vib}+ \hat{H}_\text{cav} + \hat{H}_\text{int}$. 
For the molecular chain, we consider a nearest-neighbor tight-binding model with the Peierls-type electron-phonon coupling
\begin{equation}\label{eq:H_el}
  \hat{H}_\text{el}=\sum_{n=1}^N \varepsilon_x \hat{c}_n^{\dagger}\hat{c}_n +
  \sum_{\langle mn\rangle}[-\tau+\alpha(u_{m}-u_{n})](\hat{c}_{m}^{\dagger}\hat{c}_{n} + \hat{c}_{n}^{\dagger}\hat{c}_m)
\end{equation}
Here $\hat{c}_{n}^{\dagger}$ ($\hat{c}_n$) denotes the creation (annihilation) operator of the exciton on lattice site $n$, and $\varepsilon_x$ is the on-site exciton energy. 
We model the lattice distortion as a harmonic oscillation mode with vibrational Hamiltonian $H_\text{vib}(\{u_n,\dot{u}_n\})= \frac{K}{2}u_n^2 + \frac{1}{2}M\dot{u}_n^2$ where $u_{n}$ represents the displacement of the low-frequency lattice distortion of site $n$ and $\dot{u}_{n}$ denotes its time derivative. The force constant $K$ and the molecular mass $M$ are assumed to be identical for all sites, and the characteristic vibrational frequency is given by $\omega_v=\sqrt{K/M}$.
Here $\langle mn\rangle$ denotes the nearest neighbor (i.e. $m=n\pm1$) and we set $u_{N+1}=u_{1}$ and  $\hat{c}_{N+1}=\hat{c}_{1}$ for a periodic boundary condition.

The intermolecular coupling between nearest neighbors in Eq.~\eqref{eq:H_el} is characterized by two exciton transfer mechanisms: the electronic transfer and the Peierls-type transfer (also known as Su-Schrieffer-Heeger mechanism)\cite{fratini2016transient}.
The standard electronic transfer is characterized by the nearest-neighbor transfer integral $\tau$, associated with the electronic bandwidth. 
The Peierls-type transfer mechanism is induced by the nonlocal electron-phonon coupling through the difference in displacements of neighboring molecules ($u_{m}-u_{n}$) with the coupling strength $\alpha$\cite{troisi2006charge,fratini2016transient} .
Note that we neglect the local electron-phonon coupling mechanism as induced by intramolecular vibration (Holstein-type phonon).
This model is referred to as the adiabatic Su-Schrieffer-Heeger (SSH) model, commonly used to describe conductive polymers\cite{su1979solitons,troisi2006charge}.

Now we consider the molecular chain is coupled to a single-mode cavity of photon frequency $\omega_c$ and cavity loss rate $\Gamma_c$, i.e. $\hat{H}_\text{cav} =\left(\hbar\omega_{c} -\frac{i}{2}\Gamma_c\right)\hat{a}^{\dagger}\hat{a}$, where  $\hat{a}^{\dagger}$ ($\hat{a}$) is the creation (annihilation) operator of the cavity photon mode.
For simplicity, we assume that the light-matter interactions take the form of the Tavis-Cummings model
\begin{equation}\label{eq:H_int}
\hat{H}_\text{int} = \sum_{n}g(\hat{a}^{\dagger}\hat{c}_{n} + \hat{a}\hat{c}_{n}^{\dagger}).
\end{equation}
Here we employ the rotating wave approximation (i.e. neglecting $\hat{a}^{\dagger}\hat{c}_{n}^\dagger$ and $\hat{a}\hat{c}_{n}$ terms) and the long-wavelength approximation (i.e. the coupling strength $g$ is identical for all lattice sites).
This setup corresponds to a single-layer material located near the center of a Fabry–P\'erot cavity as in Ref.~\onlinecite{mandal2023microscopic}, where the electric field is approximately uniform along the molecular chain.

With the Tavis-Cummings coupling to the electronic excitation, we focus on the cooperative photon effect in the low charge carrier density limit. 
We consider the single-exciton subspace where the charge carrier density is limited to $\rho=1/N$. 
For electronic excitons, $|G\rangle$ denotes the superposition state where all the molecular sites are in their ground state, and $|X_n\rangle=\hat{c}_n^\dagger |G\rangle$ denotes the superposition state with a single exciton on the molecular site $n$. 
For the cavity photon, we denote $|0\rangle$ as the vacuum state and $|\alpha\rangle=\hat{a}^\dagger |0\rangle$ as the single photon state. 
Within the hybrid space of excitons and photon, we denote the composite basis state as $|X_n0\rangle = |X_n\rangle \otimes |0\rangle$ and $|G\alpha\rangle = |G\rangle \otimes |\alpha\rangle$. 
Thus, in the low charge carrier limit, the total wavefunction can be written as $|\Psi(t)\rangle =\sum_{n=1}^NC_{n}(t)|X_n0\rangle+C_0(t)|G\alpha\rangle$.

\subsection{Optical conductivity and carrier mobility}
To investigate exciton transport properties under the influence of the cavity photon, we employ the linear-response Kubo formalism\cite{fetherolf2020unification,fetherolf2023conductivity}.
The key quantity in the Kubo formalism is the current autocorrelation function evaluated at thermal equilibrium
\begin{equation}\label{eq:CJJ}
    C_{JJ}(t)= \langle \hat{J}(t)\hat{J}(0)\rangle = \frac{1}{Z}\text{Tr}\{\hat{J}(t)\hat{J}(0)e^{-\beta\hat{H}}\}
\end{equation}
where the partition function is $Z=\text{Tr}\{e^{-\beta\hat{H}}\}$ and $\hat{J}(t)$ is the charge current operator. 
Here, the charge current operator can be derived from the polarization of the molecular chain\cite{ortmann2009theory} 
\begin{equation}\label{eq:8}
  \hat{J} = \frac{d\hat{P}}{dt} = -\frac{i}{\hbar}[\hat{P},\hat{H}],
\end{equation}
where the polarization of the molecular chain is given by $\hat{P} = \text{e}_0 \sum_{n}nL\hat{c}_{n}^\dagger\hat{c}_n$ and $\text{e}_0$ is the electron charge and $L$ is the lattice constant.
In the low carrier density limit, the charge carrier mobility can be obtained by
\begin{equation}\label{eq:mobility}
     \mu=\frac{\beta}{2\text{e}_0} \int_{-\infty}^{\infty} dt C_{JJ}(t)
\end{equation}
where $\beta = 1/k_BT$ and the Boltzmann constant $k_B$ and the temperature $T$.
The frequency-domain optical conductivity can be obtained by the Fourier transformation of the current autocorrelation function
\begin{equation}\label{eq:conductivity}
  \text{Re}\sigma(\omega) = \frac{1-e^{-\beta \omega}}{2N\omega}\int_{-\infty}^{\infty} dt\ 
  e^{i\omega t}C_{JJ}(t)
\end{equation}
Note that the zero-frequency component of the optical conductivity corresponds to the charge carrier mobility in Eq.~\eqref{eq:mobility}.

In the presence of the cavity, we notice that the charge current operator is contributed by two terms $\hat{J}=\hat{J}_\text{mol}+\hat{J}_\text{cav}$ where
\begin{equation}\label{eq:Jmol}
    \hat{J}_\text{mol}(t)= -\frac{iL\mathrm{e}_0}{\hbar}\sum_{\langle{mn}\rangle}[-\tau+\alpha(u_{m}-u_{n})](\hat{c}_m^{\dagger}\hat{c}_{n}-\hat{c}_{m}\hat{c}_{n}^{\dagger})
\end{equation}
and
\begin{equation}\label{eq:Jcav}
    \hat{J}_\text{cav} = -\frac{iL\mathrm{e}_0}{\hbar}\sum_{n}ng(\hat{c}_n^{\dagger}\hat{a} - \hat{c}_n\hat{a}^{\dagger})
\end{equation}
Here $\hat{J}_\text{mol}(t)$ is the current flux through the molecular chain as induced by the intermolecular coupling and $\hat{J}_\text{cav}$ is the current flux emerging from energy exchange between cavity photon and molecular excitations. 
Note that $\hat{J}_\text{mol}(t)$ is implicitly time-dependent due to the Peierls-type phonon coupling and $\hat{J}_\text{cav}$ is a time-independent operator. 
While $\hat{J}_\text{mol}(t)$ corresponds to the standard local nearest-neighbor hopping within the molecular chain, $\hat{J}_\text{cav}$ is featured by non-local coupling between exciton and photon, which can lead to long-range hopping as shown in Ref.~\onlinecite{chavez2021disorder}.
With these distinct current fluxes, the current autocorrelation can be written as $C_{JJ}(t)=C_{JJ}^\text{mol}(t)+C_{JJ}^\text{cav}(t)$ where
\begin{equation}\label{eq:CJJ_mol}
  C_{JJ}^\text{mol}(t) = \frac{1}{Z}\text{Tr}\{\mathcal{U}(0,t)\hat{J}_\text{mol}(t)\mathcal{U}(t,0)\hat{J}_\text{mol}(0)e^{-\beta \hat{H}(0)}\}
\end{equation}
\begin{equation}\label{eq:CJJ_cav}
  C_{JJ}^\text{cav}(t) = \frac{1}{Z}\text{Tr}\{\mathcal{U}(0,t)\hat{J}_\text{cav}\mathcal{U}(t,0)\hat{J}_\text{cav}e^{-\beta H(0)}\}
\end{equation}
We choose the system to be at equilibrium at $t=0$ and $\mathcal{U}(t,0)$ denotes the time-ordered propagator $\mathcal{U}(t,0) = \mathcal{T}\exp\left[-\frac{i}{\hbar} \int_{0}^{t}  \,dt'H(t')\right]$.
For simplicity, we denote the Fourier component of the current autocorrelation functions by $C^\text{mol}_{JJ}(\omega)=\int_{-\infty}^{\infty} dt\ 
  e^{i\omega t} C^\text{mol}_{JJ}(t)$ and $C^\text{cav}_{JJ}(\omega)=\int_{-\infty}^{\infty} dt\ 
  e^{i\omega t} C^\text{cav}_{JJ}(t)$.
Therefore, we can decompose the total carrier mobility into $\mu=\mu_\text{mol}+\mu_\text{cav}$, corresponding to the contributions from $\hat{J}_\text{mol}(t)$ and $\hat{J}_\text{cav}$ respectively.

\subsection{Single-particle spectral function}
We are also interested in the spectral properties of the molecular chain strongly coupled to the cavity photon.
To illustrate spectral properties, we calculate the single-particle spectral function in the frequency ($\omega$) and wave vector ($k$) space\cite{sobota2021angle,mahan_many-particle_1995},
\begin{equation}\label{eq:spectral}
A(k,\omega) = -\frac{1}{N\pi}\sum_{m,n}e^{ikL(m-n)}\text{Im}\int_{0}^{\infty} dt\ e^{i\omega t}\mathcal{G}_{mn}^{R}(t),
\end{equation}
where the retarded Green's function is given by
\begin{equation}\label{eq:Green}
i\mathcal{G}_{mn}^{R}(t) = \frac{\langle G\alpha|\hat{c}_{m}(t)\hat{c}_{n}^{\dagger}(0)e^{-\beta \hat{H}(0)}|G\alpha\rangle}{\langle G\alpha|e^{-\beta\hat{H}} |G\alpha\rangle}.
\end{equation}
Note that $|G\alpha\rangle$ is the overall ground state with one photon. 
The spectral function $A(k,\omega)$ can be interpreted as a probability distribution for excitation quasi-particle with crystal momentum $\hbar k$ and energy $\hbar\omega$.
A Lorentzian peak intensity of $A(k,\omega)$ is proportional to the quasi-particle lifetime (i.e. the inverse of the imaginary part of retarded self-energy)\cite{mahan_many-particle_1995}.
We emphasize that $A(k,\omega)$ can be reduced to the density of states by summing over the momentum dimension, $\rho(\omega) = \sum_{k}A(k,\omega)/N$.
For spectroscopic observation, the momentum-resolved spectral function can be measured by the angle-resolved photoemission spectroscopy (ARPES) and tunneling spectroscopy\cite{sobota2021angle,jang_full_2017}.

\section{Analysis without electron-phonon coupling}
\label{sec:Analysis}
\subsection{Exciton-polariton band}
Before computational simulations, it is intuitive to investigate the cavity effect on the bare electronic tight-binding model in the absence of electron-phonon coupling (i.e. $\alpha=0$).
In this limit, the single excitation matrix in Fig.~\ref{fig:schematic}(b) can be diagonalized to obtain the exciton-polariton band. 
In the absence of the cavity photon ($g=0$), the bare electronic band of the single-exciton state follows the dispersion $E_k=\varepsilon_{x}-2\tau\cos({2\pi k}/{N})$ for $k=0,\cdots,N-1$ corresponding to the eigenstates $|\Phi_{k}\rangle=\frac{1}{\sqrt{N}}\sum_{n=1}^{N}e^{i(n-1)2\pi k/N}|X_n0\rangle$. 
When coupled to the cavity photon through the interaction Hamiltonian $\hat{H}_\text{int}$, the symmetric state $|\Phi_{0}\rangle=\frac{1}{\sqrt{N}}\sum_{n=1}^{N}|X_n0\rangle$ and the cavity photon state $|G\alpha\rangle$ forms the polarironic states 
$|P_{+}\rangle=\cos\Theta|G\alpha\rangle+\sin\Theta|\Phi_{0}\rangle$ and
$|P_{-}\rangle=-\sin\Theta|G\alpha\rangle+\cos\Theta|\Phi_{0}\rangle$
where the mixing angle is defined by $\Theta=\frac{1}{2}\tan^{-1}[2g\sqrt{N}/\Delta]$ and the detuning is given by $\Delta = \hbar\omega_{c}-\varepsilon_x+2\tau$. 
The energy of the polaritonic states is given by 
\begin{equation}\label{eq:}
E_{\pm}(\Delta)=\frac{\hbar\omega_{c}+\varepsilon_x-2\tau}{2}\pm\sqrt{\frac{\Delta^{2}}{4}+g^{2}N},
\end{equation}
corresponding the upper polariton (UP) and lower polariton (LP), respectively.
Note that, since we assume the light-matter interaction with identical coupling strength to each molecule, the cavity photon couples to the symmetric state $|\Phi_{0}\rangle$ with energy $E_0=\varepsilon_x-2\tau$ (which is the lowest energy of the band). The other $N-1$ states of the bare electronic band ($E_k$ for $k=1,\cdots,N-1$) are considered optically dark to the cavity photon in the absence of electron-phonon coupling.

\subsection{Polariton effect on transport properties}
To explore the effect of polariton formation on the transport properties, we now choose the cavity photon to be resonant with the symmetric state i.e. $\Delta=\hbar\omega_{c}-\varepsilon_x+2\tau=0$, so that the energy splitting is $E_{\pm}=\varepsilon_{x}-2\tau \pm\sqrt{N}g$.
Interestingly, we find that the polariton states do not contribute to the current autocorrelation function of the molecular chain $C_{JJ}^\text{mol}(t)$ since both the symmetric state and the photon state does not contribute to the current flux through the molecular chain (i.e. $\hat{J}_\text{mol}(0)|\phi_0\rangle=\hat{J}_\text{mol}(0)|G\alpha\rangle=0$), so that $\langle P_\pm|\hat{J}_\text{mol}(t)\hat{J}_\text{mol}(0)|P_\pm\rangle=0$. 
Thus, the current flux $C_{JJ}^\text{mol}(t)$ is solely contributed by the dark states, more explicitly $\hat{J}_\text{mol}(0)|\phi_k\rangle=2\tau L\sin(\frac{2\pi k}{N})|\phi_k\rangle$ for $k=1,\cdots,N-1$. 
However, for the ensemble average in Eq.~\eqref{eq:CJJ_mol}, the weight of the lower polariton state $e^{-\beta (\varepsilon_{x}-2\tau-\sqrt{N}g)}/Z$ grows as $\sqrt{N}g$ increases, thereby the zero contribution term dominates.
Consequently, we find that, in the absence of the electron-phonon coupling, forming polaritonic states decreases $C_{JJ}^\text{mol}(t)$, thereby reducing the mobility $\mu_\text{mol}$.

Next, we focus on the current operator through the cavity photon $\hat{J}_\text{cav}$.
With the analytical expressions for the polariton states, we can estimate the autocorrelation function by its LP contribution (see Appendix~\ref{sec:appendix})
\begin{equation}\label{eq:CJJ_cav_analytical}
    C_{JJ}^\text{cav}(\omega)\approx\frac{\text{e}_0^2g^2L^2}{2\hbar}N^3\left[\frac{\Gamma_c}{(\hbar\omega-2\sqrt{N}g)^2+\Gamma_c^2}-\frac{1}{2\pi^2} \frac{\Gamma_c/2}{(\hbar\omega-\sqrt{N}g)^2+\Gamma_c^2/4}\right]
\end{equation}
which predicts a two-peak structure: a higher peak at $\hbar\omega=E_+-E_-=2\sqrt{N}g$ and a lower peak at $\hbar\omega=E_0-E_-=\sqrt{N}g$. 
This predicted structure gives us an approximate picture of the optical conductivity contributed by the cavity current flux in the absence of electron-phonon coupling. 
However, we note that the zero-frequency component of the LP contribution $C_{JJ}^\text{cav}(\omega\rightarrow0)\propto\frac{\text{e}_0^2L^2N^2}{2\hbar}$ does not increase with $g$, implying that the mobility through cavity photon $\mu_\text{cav}$ saturate to a constant as $g$ increases.

\section{Simulation Results}\label{sec:results}
\begin{figure}[ht]
  \centering
  \includegraphics[width=0.8\linewidth]{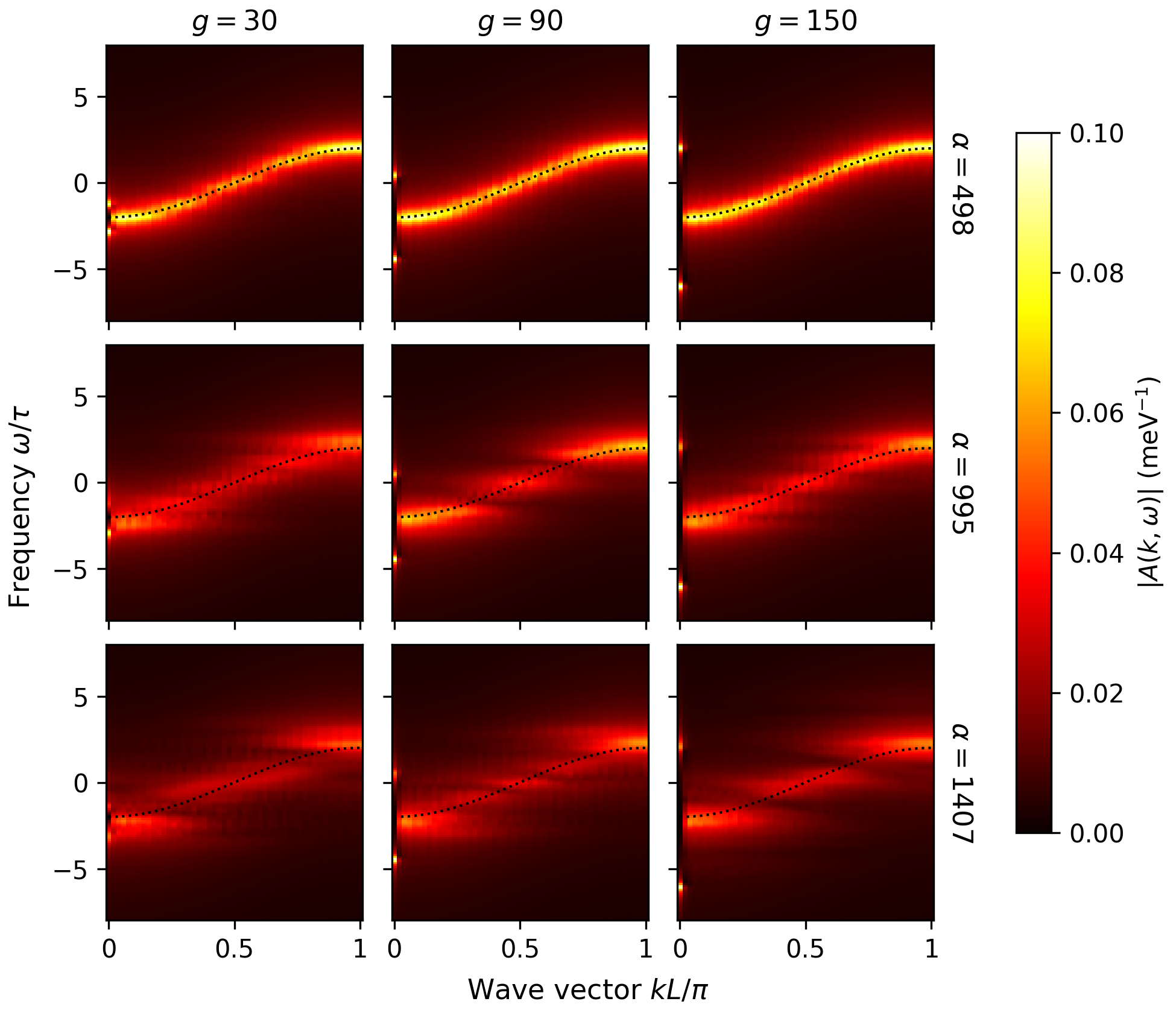}
  \caption{Momentum-resolved spectral function $A(k,\omega)$ with varying electron-phonon coupling strength $\alpha = 498,995,1407$ (in the unit of $\text{cm}^{-1}/$\AA), and light-matter coupling strength $g = 30, 90, 150$ (in the unit of $\text{cm}^{-1}$).
  For all parameters, we observe the upper and lower polariton peaks ($\omega=E_\pm$) at $k=0$ and a broadened version of the electronic band for $k>0$. The dashed black line indicates the bare electronic band. While the electronic band is not affected by the cavity phonon (except at $k=0$), the polariton peaks are lowered by dynamic disorder and broadened in the $k$ dimension, especially when $g$ is small.}
  \label{fig:SpectralFunction}
\end{figure}
\begin{figure}[ht]
  \centering
  \includegraphics[width=0.8\linewidth]{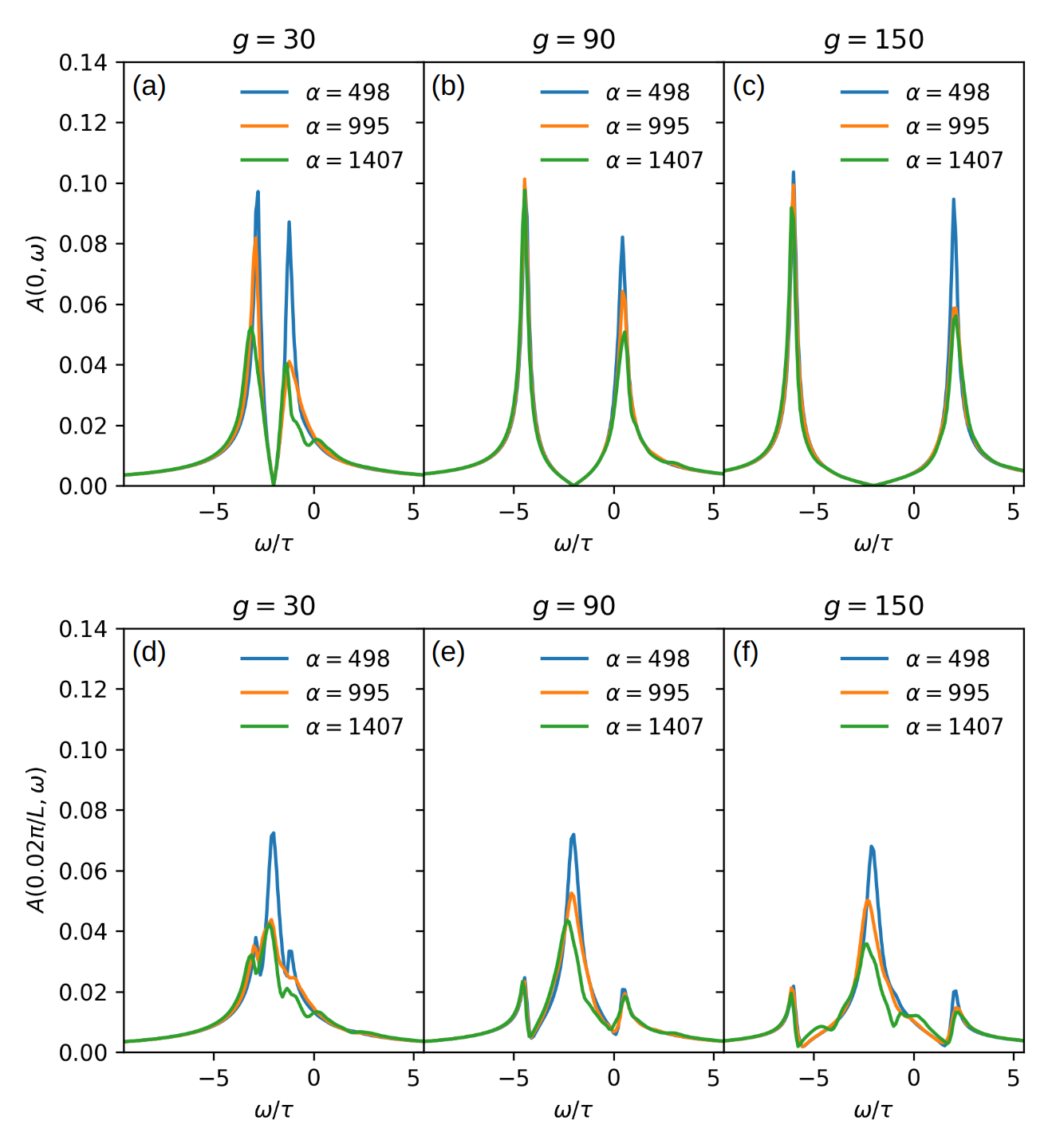}
  \caption{The spectral functions $A(k=0,\omega)$ for (a) $g=30$, (b) $g=90$, and (c) $g=150$ (in the unit of $\text{cm}^{-1}$) exhibit the UP/LP peaks with decreasing heights as $\alpha$ increases. The UP peaks are suppressed more than the LP in the presence of dynamic disorder. In the lower panels, $A(k=0.02\pi/L,\omega)$ for (d) $g=30$, (e) $g=90$, and (f) $g=150$ (in the unit of $\text{cm}^{-1}$) shows three peaks corresponding to the electronic band (middle) and the UP and LP side peaks (right and left). The polariton side peaks appear only when electron-phonon coupling is present, and can be suppressed as $\alpha$ increases. }
  \label{fig:A0w_cut}
\end{figure}

\subsection{Parameters and Simulation details}
Following Ref.~\onlinecite{troisi2006charge}, we consider the parameter for typical organic molecular crystals with elastic constant $K = 14500\ \text{amu}\cdot\text{ps}^{-2}$\ \text{and}\ $M=250\ \text{amu}$ (pentacene) so that Peierls phonon frequency is $\omega_v=40.4\ \text{cm}^{-1}$.
The transfer integral is $\tau=300\ \text{cm}^{-1}$ and the typical lattice constant is $L=4$ \AA. We set the temperature $T = 300K$ and the lattice size $N= 64$. The electron-phonon coupling strength is within the range $\alpha = 500-1500 \ \text{cm}^{-1}$/\AA. 
To illustrate the interplay between the intermolecular Peierls phonon and the optical cavity photon, we choose the cavity photon frequency to be resonant with the symmetric state $|\Phi_0\rangle$, i.e. $\hbar\omega_c=\varepsilon_x-2\tau$, and the cavity decay rate is $\Gamma_c=600\ \text{cm}^{-1}$. The light-matter coupling strength is within the range $g=30-150\ \text{cm}^{-1}$.

To numerically evaluate the current autocorrelation functions in Eq.~\eqref{eq:CJJ_mol} and \eqref{eq:CJJ_cav}, we employ quasiclassical dynamics simulations where the electronic and photonic degrees of freedom follow fully quantum dynamics $i\hbar\frac{\partial}{\partial t}|\Psi(t)\rangle=\hat{H}|\Psi(t)\rangle$ while the lattice distortion modes are assumed to follow classical dynamics $M\ddot{u}_{n} = -Ku_{n}$.
Here we neglect the non-adiabatic force $-\frac{\partial}{\partial u_{n}}\langle\Psi(t)|\hat{H}_\text{el}(t)|\Psi(t)\rangle$ on the lattice distortion since the phonon frequency is small compared to $\tau$\cite{troisi2006charge,ciuchi2011transient}.
To evaluate the current autocorrelation function, we integrate the Schrodinger equation and evaluate $\mathcal{U}(t,0)|\Psi\rangle$ and $\mathcal{U}(t,0)\hat{J}|\Psi\rangle$ where $\hat{J}$ can be either $\hat{J}_\text{mol}$ or $\hat{J}_\text{cav}$.
The initial states $|\Psi\rangle$ are the eigenstates of the initial Hamiltonain $\hat{H}(0)$.
We sample the initial phonon displacement $u_n(0)$ and $\dot{u}_n(0)$ following the equilibrium distribution of the phonon mode $e^{-\beta H_\text{vib}(\{u_n,\dot{u}_n\})}$ and average over $1000$ classical trajectories.
The time step is set to $dt = 0.0025\ \text{fs}$ to ensure numerical stability, and the Schrodinger equation is integrated by 4-th order Runge-Kutta method with a total time of $t_\text{max}=0.3\ \text{ps}$. 



\subsection{Momentum-resolved spectral functions}\label{subsec:spectral}

We first illustrate the momentum-resolved spectral function $A(k,\omega)$ with various electron-phonon coupling $\alpha$ and light-matter coupling strength $g$. 
In Fig.~\ref{fig:SpectralFunction}, we observe the upper/lower polaritons exhibit the Rabi splitting $2\sqrt{N}g$ at $kL=0$ for all parameters, as explained in Sec.~\ref{sec:Analysis}, and the bare electronic band is depicted by the black dotted line.
To observe the phonon influence on the polaritonic states in detail, Fig.~\ref{fig:A0w_cut} (a--c) shows the spectral function at $kL=0$, i.e. $A(0,\omega)$ for different light matter coupling strength $g=30, 90, 150~\text{cm}^{-1}$. 
The overall trend shows that increasing electron-phonon coupling strength $\alpha$ reduces the peak intensity for both UP and LP. 
We notice that, for a weak electron-phonon coupling ($\alpha=498\ \text{cm}^{-1}/\text{\AA}$, blue lines), the UP peak is slightly lower than the LP peak. 
However, for stronger electron-phonon coupling,  the UP peak intensity is significantly suppressed, which corresponds to a shorter lifetime of the UP state than of the LP state as induced by dynamic disorder (electron-phonon coupling)\cite{park_polariton_2022}.
Furthermore, we find that, the shape of the UP peak is disturbed by dynamic disorder, especially when $g$ is small, while the LP peak retains the Lorentzian shape. 
This can be explained by the fact that the UP state is within the electronic band ($-2\tau<\omega<2\tau$) and mixes with the dark states due to electron-phonon coupling. This observation agrees with previous studies\cite{coles2011vibrationally,coles2013imaging,tichauer2022identifying}, in which molecular vibrational modes play profound roles in relaxation dynamics of polaritonic states.

In the regime $0<kL<\pi$ of Fig.~\ref{fig:SpectralFunction}, we find that the spectral function follows a broadened version of the non-interacting dispersion relation $E_k = \varepsilon_x -2\tau\cos(kL)$.
Similar to Ref. \onlinecite{fetherolf2020unification}, the spectral function near the band center $kL=\pi/2$ and edges $kL=0,\pi$ is broadened in the $k$ dimension when the electron-phonon coupling is strong ($\alpha=1407\ \text{cm}^{-1}/\text{\AA}$, bottom panels).  
The broadened band implies that phonon-dressed quasiparticles can make direct transitions for a given $k$, suggesting lower zero-frequency components and peak structures in optical conductivity (which we discuss in Sec.~\ref{subsec:conductivity}). 
In addition to this broadening effect, we find the band edges are slightly expanded in the $\omega$ dimension with increasing $\alpha$.
These phonon-dressed electronic bands are not affected by the light-matter coupling since the cavity photon frequency is much larger than the Peierls phonon mode ($\omega_v\ll\omega_c$). 
That being said, we note that the spectral function at the UP and LP energies is also broadened in the $k$ dimension due to electron-phonon coupling. 
In Fig.~\ref{fig:A0w_cut} (d--f),  we plot the spectral function at $kL/\pi=0.02$ and observe three peaks corresponding to the UP/LP ($\omega=E_{\pm}=-2\tau\pm\sqrt{N}g$) and the phonon-dressed electronic band near $\omega=-2\tau$. 
The separation of the polariton side peaks is proportional to $g$ and the LP peak is not suppressed by dynamic disorder.
This three-peak structure implies that a direct transition is enabled between the polaritons and the phonon-dressed electrons, thereby their corresponding conductivity peaks should shift as $g$ increases (which we discuss in Sec.~\ref{subsec:conductivity}).

\subsection{Charge carrier mobility}\label{subsec:mobility}
\begin{figure}[ht]
    \centering
    \includegraphics[width=\linewidth]{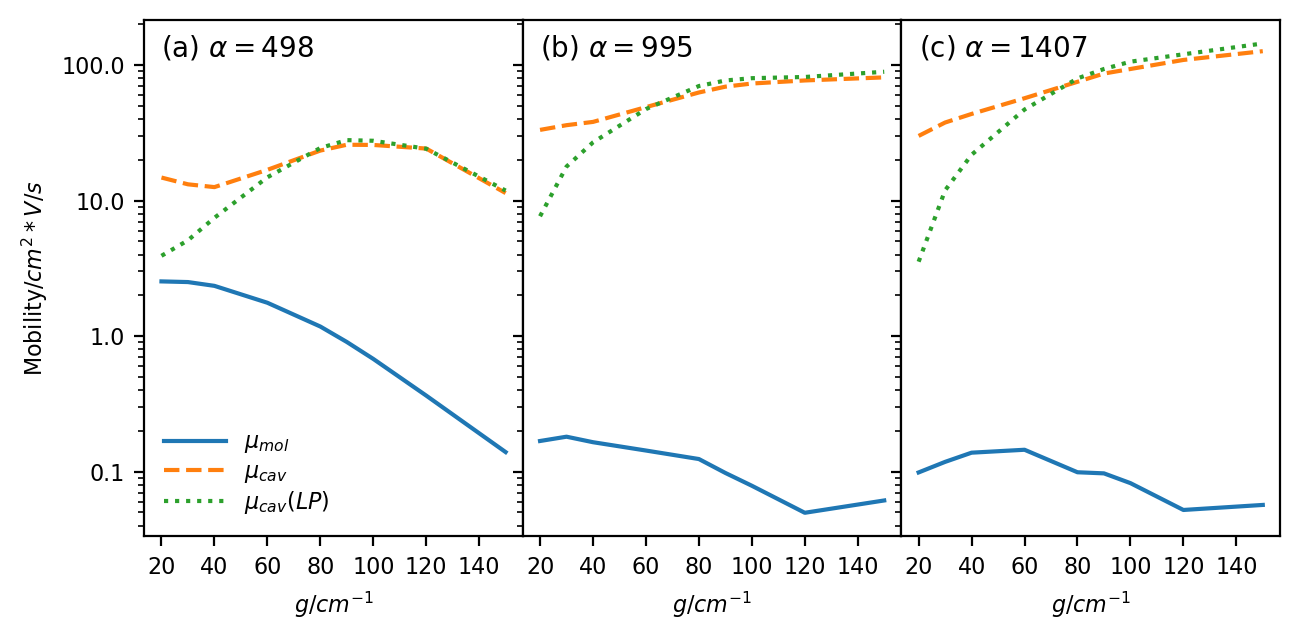}
    \caption{Charge carrier mobilities, $\mu_\text{mol}$ (solid line), $\mu_\text{cav}$ (dotted line) and its LP contribution $\mu_\text{cav}(\text{LP})$ (dashed line), are plotted as a function of light-matter coupling strength $g$ for (a) $\alpha = 498$, (b) $\alpha = 995$, and (c) $\alpha = 1407$ (in the unit of $\text{cm}^{-1}/$\AA). In general, $\mu_\text{cav}>\mu_\text{mol}$ and $\mu_\text{mol}$ is suppressed as $g$ increases. The difference between $\mu_\text{cav}$ and $\mu_\text{cav}(\text{LP})$ is attributed to the optical dark states, which contribute to the mobility for small $g$. For large $g$, the LP contribution of the mobility is enhanced as $\alpha$ increases, showing vibration-assisted transport. }\label{fig:dc_g} 
\end{figure}
Now we turn our attention to the charge carrier mobility calculated by Eq.~\eqref{eq:mobility}. Fig.~\ref{fig:dc_g} exhibits the corresponding mobility ($\mu_\text{mol}$ and $\mu_\text{cav}$) as a function of the light-matter coupling strength ($g$). 
Interestingly, we find that, for all electron-phonon coupling regime, the mobility through the molecular chain ($\mu_\text{mol}$) tends to decrease as $g$ increases. As discussed in Sec.~\ref{sec:Analysis}, this decrease can be attributed to the increasing weight of the LP state in the partition function $Z=e^{-\beta E_-}+e^{-\beta E_+}+\sum_{k=1}^{N-1}e^{-\beta E_k}$, which does not contrite to the current flux through the molecular chain. Thus, the formation of the polaritons substantially suppresses $\mu_\text{mol}$, especially in the weak dynamic disorder regime ($\alpha = 498\ \text{cm}^{-1}/\text{\AA}$, see the solid line in Fig.~\ref{fig:dc_g}(a)).
On the other hand, in the strong dynamic disorder regime ($\alpha = 995, 1407\ \text{cm}^{-1}/\text{\AA}$), $\mu_\text{mol}$ is overall suppressed by dynamic disorder.

Next, we observe that the charge carrier mobility contributed by the cavity current flux ($\mu_\text{cav}$, dashed lines) dominates the mobility for all parameters we consider, especially in the strong electron-phonon coupling regime. 
Importantly, by comparing Fig.~\ref{fig:dc_g} (a--c), we find that increasing electron-phonon coupling enhances $\mu_\text{cav}$. To understand this enhancement, we compare the total $\mu_\text{cav}$ and the contribution from the lower polariton state $\mu_\text{cav}\text{(LP)}$.
In the weak light-matter coupling regime ($g<60\ \text{cm}^{-1}$ or $g/\tau<0.2$), we notice that $\mu_\text{cav}>\mu_\text{cav}\text{(LP)}$, suggesting that the dark states significantly contribute to the current flux  through the cavity and this contribution is more profound when $\alpha$ is large (see panel (c)). This dark state contribution to $\mu_\text{cav}$ arises from the delocalization characteristics of the current operator---given $\hat{J}_\text{cav}\propto \sum_n ng(\hat{c}_n^\dagger\hat{a}-\hat{c}_n\hat{a}^\dagger)$, the current between site $n$ and photon has an explicit dependence on $n$.
Similar disorder-induced transport phenomena are discussed by Celardo $et\ al$ \cite{chavez2021disorder}, in which the cavity photon mode enables an effective long-range hopping via the dark states. 

By contrast, in the strong light-matter coupling regime ($g>120\ \text{cm}^{-1}$ or $g/\tau>0.4$), $\mu_\text{cav}$ is mostly contributed by the LP state, and the dark states turn out to reduce charge carrier transport (i.e. $\mu_\text{cav}<\mu_\text{cav}\text{(LP)}$) as opposed to enhancement for small $g$.
In particular, Fig.~\ref{fig:dc_g}(a) shows that $\mu_\text{cav}\text{(LP)}$ is suppressed as $g$ increases, which can be attributed to the increasing weight of the LP state in the partition function $Z$ (showing a similar trend as $\mu_\text{mol}$). 
In the presence of strong dynamic disorder, however, $\mu_\text{cav}\text{(LP)}$ is enhanced and plateaus as $g$ increases.
This enhancement results from a combination of the effective long-range hopping mechanism mediated by the cavity photon and vibration-assisted scattering as induced by electron-phonon coupling\cite{coles2011vibrationally, coles2013imaging, tichauer2022identifying, litinskaya2004fast,krupp2024quantum}.

%


\subsection{Optical conductivity}\label{subsec:conductivity}

\begin{figure}[ht]
    \centering
    \includegraphics[width=\linewidth]{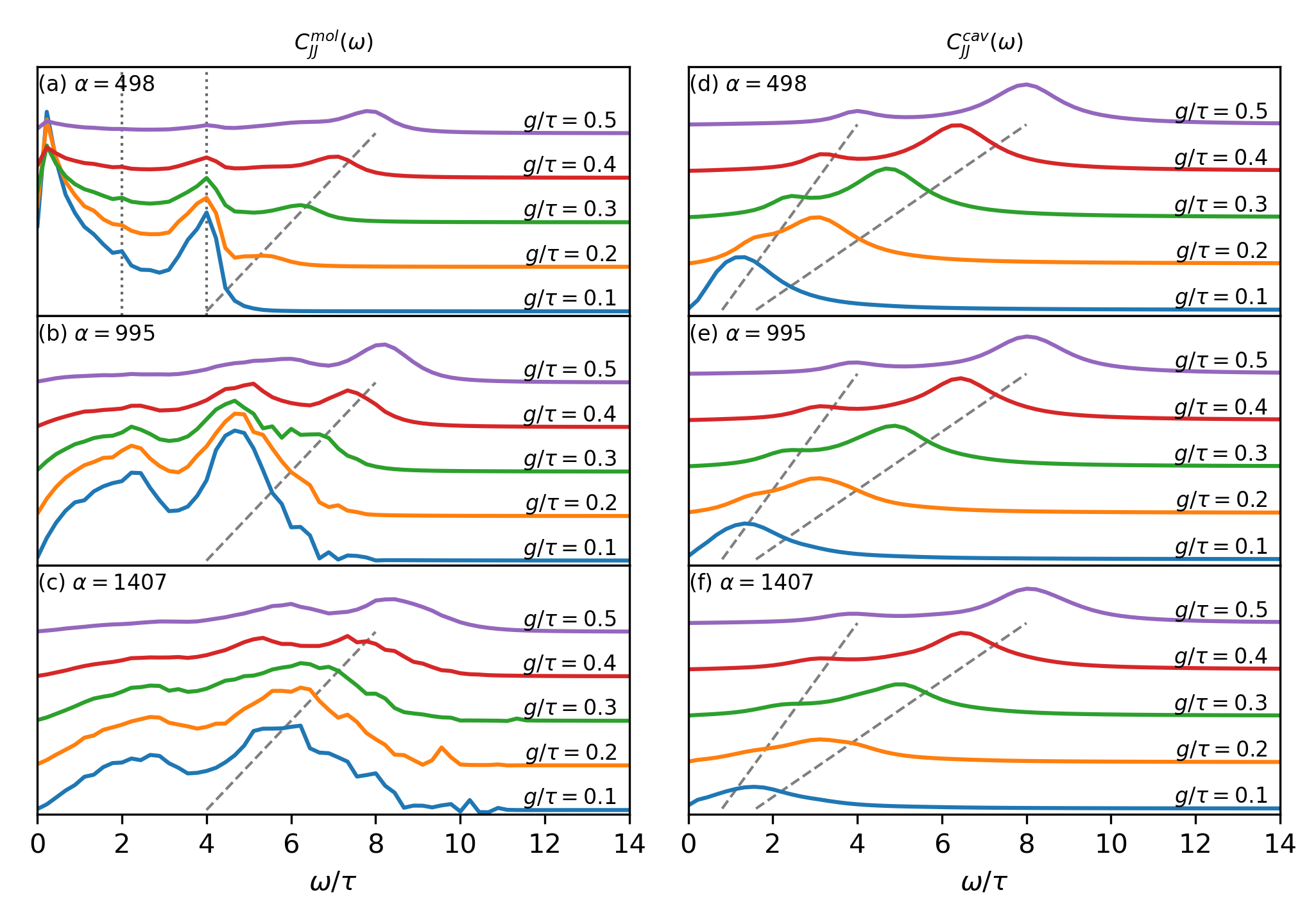}
    \caption{Qualitative trend of $C_{JJ}^\text{mol}(\omega)$ and $C_{JJ}^\text{cav}(\omega)$ with increasing $g$ for dynamical disorder $\alpha = 498,\ 995,\ 1407$ (in the unit of $\text{cm}^{-1}/$\AA). To show the linear scaling of the polariton side peak position, we shift the data by an offset proportional to $g$ and scale $C_{JJ}^\text{cav}(\omega)$ by $1/g^3$ for clarity. The vertical dotted lines in (a) indicate the conductivity peaks at $\omega/\tau=2,4$. The dashed lines denote the side peaks shifting linearly with $g$.}
    \label{fig:ac_1}
\end{figure} 
We further explore the optical conductivity in terms of the separated contributions of the frequency-resolved current autocorrelation functions i.e. $\text{Re}\sigma(\omega) = \frac{1-e^{-\beta \omega}}{2N\omega}[C_{JJ}^\text{mol}(\omega)+C_{JJ}^\text{cav}(\omega)]$. 
To observe the qualitative change with varying light-matter coupling strength ($g$), we plot $C_{JJ}^\text{mol}(\omega)$ and $C_{JJ}^\text{cav}(\omega)$ with a $y$-axis offset proportional to $g$ in Fig.~\ref{fig:ac_1}.
For weak dynamic disorder $\alpha=498\ \text{cm}^{-1}/$\AA~in Fig.~\ref{fig:ac_1}(a), two phonon-induced conductivity peaks are observed around $\omega/\tau=2,4$ , which agree with previous studies without coupling to a cavity\cite{cataudella2011transport, fetherolf2020unification}. The polariton side peak emerges from the coupling between the polaritonic states and the phonon-dressed electronic band, which shifts linearly with $g$. 
Note that this side peak is derived from the direct transition between the three peaks of $A(k,\omega)$ at $k=0.02\pi/L$ in Fig.~\ref{fig:SpectralFunction}(d)--(f). 
However, the polariton side peaks are not obvious when $g/\tau$ is small. 
With increasing dynamic disorder, the phonon-induced peaks are broadened and blue-shifted, corresponding to the expanded band edges in the spectral function as shown in the bottom panels of Fig.~.\ref{fig:SpectralFunction}. 
For strong dynamic disorder $\alpha=1407\ \text{cm}^{-1}/$\AA in Fig.~\ref{fig:ac_1}(c), the polariton side peak is not obvious and overlapped with the broadened phonon-induced peaks when $g/\tau$ is small. 
We notice that $C_{JJ}^\text{mol}(\omega)$ is generally suppressed at $\omega\approx0$, which also agrees with previous results. 

On the other hand, the current autocorrelation functions through the cavity $C_{JJ}^\text{cav}(\omega)$ exhibit a two-Lorentzian structure (as predicted by the approximated spectrum in Eq.~\eqref{eq:CJJ_cav_analytical}), instead of phonon-induced peaks.
Here, we find the Lorentzian peak at $\omega = 2\sqrt{N}g$ corresponds to the transition between the LP and UP states (i.e. the first term in Eq.~\eqref{eq:CJJ_cav_analytical}) and the other Lorentzian peak at $\omega = \sqrt{N}g$ corresponds to the transition from the LP state to the dark states (i.e. the seond term in Eq.~\eqref{eq:CJJ_cav_analytical}).

\begin{figure}[h]
    \centering
    \includegraphics[width=0.5\linewidth]{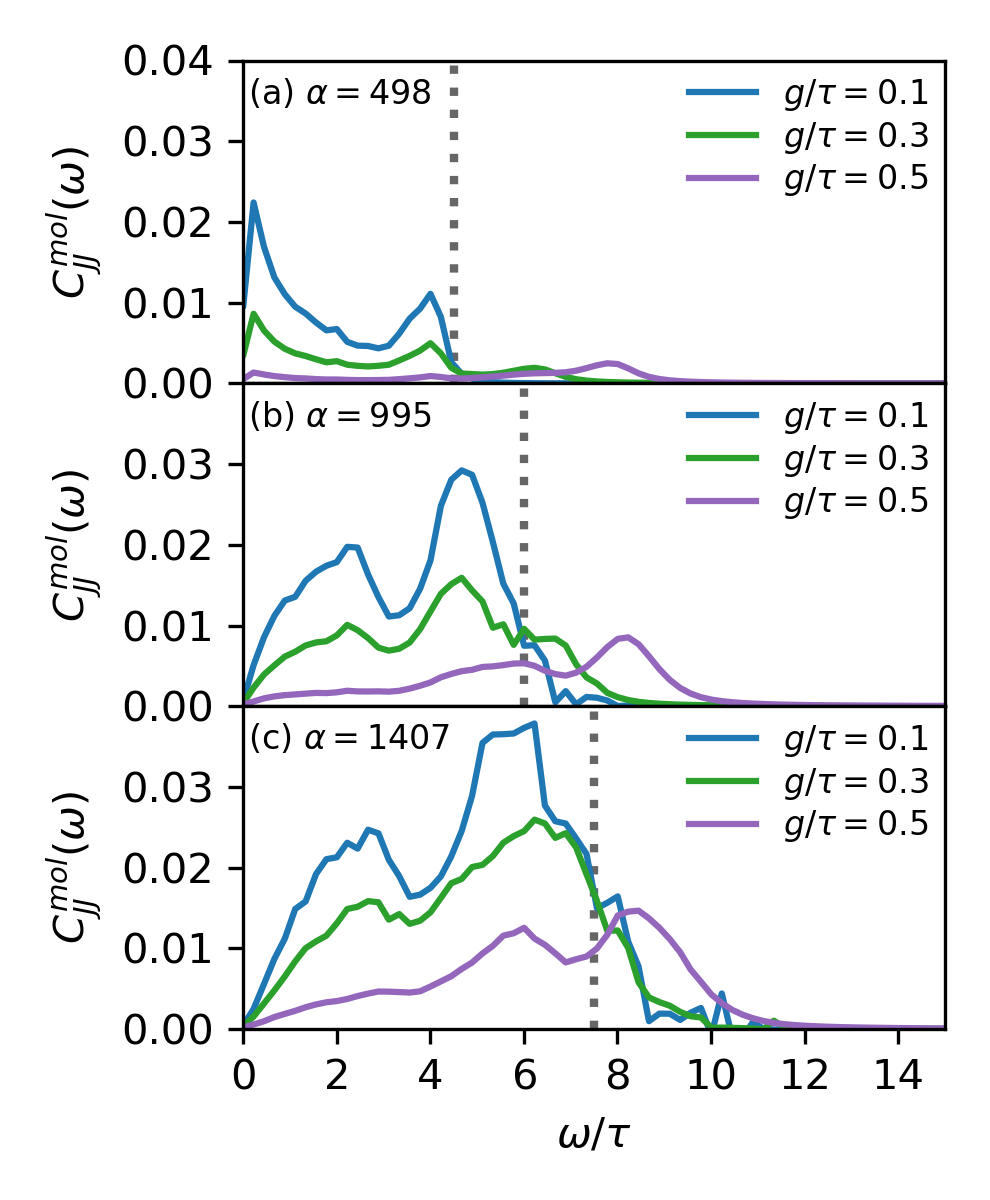}
    \caption{Frequency-domain current autocorrelation function $C_{JJ}^\text{mol}(\omega)$ at different light-matter coupling strength $g/\tau = 0.1,\ 0.3,\ 0.5$ for (a) $\alpha=498$, (b) $\alpha=995$ , and (c) $\alpha=1407$ (in the unit of $\text{cm}^{-1}/\text{\AA}$). The vertical dotted lines indicate the effective bandwidth in the presence of dynamic disorder as shown in Fig.~\ref{fig:SpectralFunction}. The phonon-induced peaks are observed within the bandwidth and reduced as $g$ increases. The polariton side peak becomes visible outside of the bandwidth and grows as $g$ increases.}
    \label{fig:ac_2}
\end{figure}

For quantitative analysis, we compare $C_{JJ}^\text{mol}(\omega)$ for $g=30, 90, 150\ \text{cm}^{-1}$ in Fig.~\ref{fig:ac_2}(a--c). 
Within the effective bandwidth (the gray dotted lines estimated by the expansion of the band edge in Fig.~\ref{fig:SpectralFunction}), $C_{JJ}^\text{mol}(\omega)$ is dominated by the phonon-induced conductivity peaks.
While the peak positions are not influenced by the light-matter coupling, the height of the conductivity peaks is overall suppressed as $g$ increases, which coincides with the suppression of the mobility $\mu_\text{mol}$. 
This suppression is also attributed to the increase in the LP state weight in the partition function.
Outside of the effective bandwidth, the polariton side peaks emerge as $g/\tau$ increases. Interestingly, the height of side peaks is enhanced as $\alpha$ increases.
This enhancement implies that the electron-phonon coupling can assist exciton transport via polaritonic states.

\begin{figure}[h]
    \centering
    \includegraphics[width=0.5\linewidth]{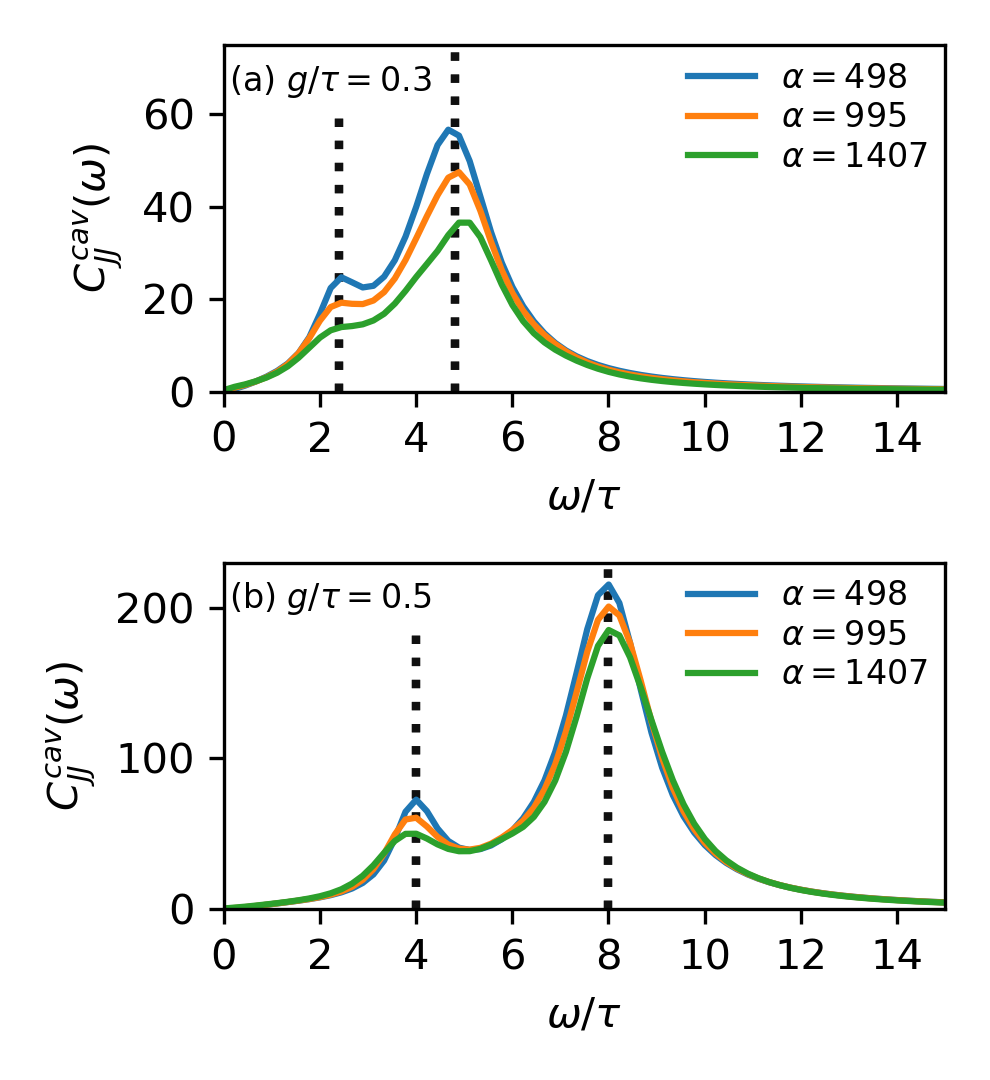}
    \caption{Frequency-domain current autocorrelation function $C_{JJ}^\text{cav}(\omega)$ at different dynamical disorder $\alpha = 498, 995 , 1407\ \text{cm}^{-1}/\text{\AA}$ for (a) $g/\tau = 0.3$ and (b) $g/\tau = 0.5$. The back dotted lines indicate $\hbar\omega=\sqrt{N}g, 2\sqrt{N}g$, repectively. Note that increasing $\alpha$ leads to broadening of the Lorentzian width and enhancement of $C_{JJ}^\text{cav}(\omega=0)$.}
    \label{fig:ac_3}
\end{figure}

In Fig.~\ref{fig:ac_3}, we show that $C_{JJ}^\text{cav}(\omega)$ shows two Lorentzian peaks as predicted by Eq.~\eqref{eq:CJJ_cav_analytical}. 
The peak width is approximately $\frac{\Gamma_c}{2}=300\ \text{cm}^{-1}=\tau$ for the peak at $\hbar\omega=2\sqrt{N}g$ and $\frac{\Gamma_c}{4}=150\ \text{cm}^{-1}=\frac{\tau}{2}$ for the peak at $\hbar\omega=\sqrt{N}g$. 
We further notice that, as $\alpha$ increases, while the peak position does not change, the peak height is reduced and the width is slightly broadened.
Thus, their zero-frequency component is enhanced due to broadening, which explains the $\mu_\text{cav}$ mobility enhancement as observed in Fig.~\ref{fig:dc_g}.
As a final note, we observe that the cavity-mediated current flux dominates the overall exciton-polariton transport, i.e. $\mu_\text{cav}\gg\mu_\text{mol}$ and $C_{JJ}^\text{cav}(\omega)\gg C_{JJ}^\text{mol}(\omega)$.

\section{Conclusion}\label{sec:conclude}
In this paper, we explore the spectral and transport properties of a molecular chain inside an optical cavity using quasi-classical dynamics. We specifically focus on the interplay between exciton-polariton and dynamic disorder as induced by Peierls-type phonon modes.
The spectral function exhibits the formation of exciton-polariton near $k=0$ and broadening induced by dynamic disorder.
We demonstrate the dissection of current flux into its standard exciton transport component and a cavity photon contribution. Notably, the exciton transport is suppressed by light-matter coupling, whereas cavity photon-mediated transport is enhanced and further assisted by electron-phonon coupling. 
We interpret this enhancement as a combination of effective long-range hopping and vibration-assisted scattering.
This observation is further confirmed by frequency-resolved conductivity.
The enhancement of carrier mobility achieved through light-matter interactions and electron-phonon coupling opens up new avenues for future material design.

Looking forward, several immediate questions need to be addressed. First, we aim to extend our current theoretical framework to include multiple cavity modes. In this scenario, multimode polaritons will form through coupling to more states of electronic bands (rather than solely the symmetric state), which will further complicate the spectral function and optical conductivity.
Second, the dependence on temperature and phonon frequency must be explored. Investigating the scaling of mobility as a function of temperature will offer a direct connection with experimental observations\cite{coles2011vibrationally}.
Finally, our current simulation method relies on quasi-classical dynamics, which neglects the Ehrenfest force on the phonon modes\cite{wang_mixed_2011}. Also, the system Hamiltonian currently includes only single excitation without incorporating many-body interactions\cite{ghosh_mean-field_2025}. Further methodological improvements are therefore necessary and will reveal a more insightful understanding of exciton-polariton transport in disordered molecular systems.

\begin{acknowledgments}
We appreciate useful suggestions from Professor Boldizsar Janko. This research is supported by the University of Notre Dame.
\end{acknowledgments}

\appendix

\section{Derivation of Eq.~\eqref{eq:CJJ_cav_analytical}}\label{sec:appendix}
Given the current operator $\hat{J}_\text{cav}=-\frac{iL\text{e}_0g}{\hbar}\sum_n n(|X_n0\rangle\langle G\alpha|-|G\alpha\rangle\langle X_n0|)$, we can evaluate the following matrix elements in the basis of the polariton states $|P_\pm\rangle$ and the dark states $|\Phi_k\rangle$ for $k=1,\cdots,N-1$. For the polariton states, we have $\langle P_+|\hat{J}_\text{cav}|P_+\rangle=\langle P_-|\hat{J}_\text{cav}|P_-\rangle=0$ and $\langle P_-|\hat{J}_\text{cav}|P_+\rangle=-\langle P_+|\hat{J}_\text{cav}|P_-\rangle=\frac{N(N+1)L}{2\sqrt{N}}$. 
For the dark state, we have $\langle \Phi_k|\hat{J}_\text{cav}|\Phi_{k'}\rangle=0$, and the cross terms are 
$\langle P_+|\hat{J}_\text{cav}|\Phi_k\rangle=\frac{LR_k}{\sqrt{N}}\cos\Theta$ and $\langle P_-|\hat{J}_\text{cav}|\Phi_k\rangle=-\frac{LR_k}{\sqrt{N}}\sin\Theta$.
Here we define $R_k=\sum_nne^{i(n-1)2\pi k/N}$, which can be evaluated using the finite arithmetic-geometric series. 
Since the weight of the LP contribution $e^{-\beta E_-}$ dominates the ensemble average as $\sqrt{N}g$ is large, we approximate the current autocorrelation by the contribution of the LP state 
\begin{equation}
    C_{JJ}^\text{cav}(t)\approx\langle P_-|e^{i\hat{H}t/\hbar}\hat{J}_\text{cav}e^{-i\hat{H}t/\hbar}\hat{J}_\text{cav}|P_-\rangle
\end{equation}
More explicitly, we have
\begin{equation}
    C_{JJ}^\text{cav}(t)\approx \frac{\text{e}_0^2g^2L^2}{\hbar^2}\left[\frac{N(N-1)^2}{4}e^{-i(E_+-E_--i\Gamma_c)t/\hbar}-\frac{1}{N}\sin^2\Theta\sum_{k=1}^{N-1}|R_k|^2e^{-i(E_k-E_--\frac{i}{2}\Gamma_c)t/\hbar}\right]
\end{equation}
and 
\begin{equation}
|R_k|^2=\frac{N^2}{4\sin^2(\pi k/N)}
\end{equation}
Next, we notice that, for large $N$, $|R_k|^2$ behaves as a delta function and is dominated by $|R_1|^2\approx N^4/4\pi^2$. Thus, we keep only the $|R_1|^2$ term of the summation and further approximate ($N\approx N-1$)
\begin{equation}
    C_{JJ}^\text{cav}(t)\approx\frac{\text{e}_0^2g^2L^2}{\hbar^2}\left[\frac{N^3}{4}e^{-i(E_+-E_--i\Gamma_c)t/\hbar}-\frac{N^3}{4\pi^2}\sin^2\Theta e^{-i(E_1-E_--\frac{i}{2}\Gamma_c)t/\hbar}\right]
\end{equation}
When $\Delta=0$, we have $\sin\Theta=\frac{1}{\sqrt{2}}$ and take the Fourier transform yields
\begin{equation}
    C_{JJ}^\text{cav}(\omega)\approx\frac{\text{e}_0^2g^2L^2}{2\hbar}N^3\left[\frac{\Gamma_c}{(\hbar\omega-(E_+-E-))^2+\Gamma_c^2}-\frac{1}{2\pi^2} \frac{\Gamma_c/2}{(\hbar\omega-(E_1-E_-))^2+\Gamma_c^2/4}\right]
\end{equation}
Finally, by taking $E_1\approx E_0=\varepsilon_x-2\tau$ (for large $N$) i.e. $E_1-E_-\approx E_0-E_-=\sqrt{N}g$ and $E_+-E_-=2\sqrt{N}g$, we can derive Eq.~\eqref{eq:CJJ_cav_analytical}.

\bibliography{exciton_polariton}
\end{document}